# Layer-Specific Hole Concentrations in $Bi_2Sr_2(Y_{1-x}Ca_x)Cu_2O_{8+\delta}$ as Probed by XANES Spectroscopy and Coulometric Redox Analysis

M. Karppinen

*Materials and Structures Laboratory, Tokyo Institute of Technology, Yokohama 226-8503, Japan,*
*and Laboratory of Inorganic and Analytical Chemistry, Helsinki University of Technology,*
*FIN-02015 Espoo, Finland*

M. Kotiranta

*Laboratory of Inorganic and Analytical Chemistry, Helsinki University of Technology,*
*FIN-02015 Espoo, Finland*

T. Nakane and H. Yamauchi

*Materials and Structures Laboratory, Tokyo Institute of Technology, Yokohama 226-8503, Japan*

S. C. Chang and R. S. Liu

*Department of Chemistry, National Taiwan University, Taipei, Taiwan, Republic of China*

J. M. Chen

*Synchrotron Radiation Research Center, Hsinchu, Taiwan, Republic of China*

Induction of holes not only in the superconductive $CuO_2$ plane but also in the $Bi_2O_{2+\delta}$ charge reservoir of the $Bi_2Sr_2(Y_{1-x}Ca_x)Cu_2O_{8+\delta}$ superconductor upon $Ca^{II}$-for-$Y^{III}$ substitution is evidenced by means of two independent techniques, *i.e.*, high-resolution x-ray-absorption near-edge structure (XANES) spectroscopy measurements and coulometric redox titrations. The absolute values derived for the $CuO_2$-plane hole concentration from the Cu $L_{2,3}$-edge XANES spectra are in good agreement with those obtained from the coulometric redox analysis. The $CuO_2$-plane hole concentration is found to increase from 0.03 to 0.14 concomitantly with the increase in the $BiO_{1+\delta/2}$-layer hole concentration from 0.00 to 0.13 as the Ca-substitution level, $x$, increases from 0 to 1. The threshold $CuO_2$-plane hole concentration for the appearance of superconductivity is determined at 0.06, while the highest $T_c$ is obtained at the hole concentration of 0.12. In the O $K$-edge XANES spectrum, the increases in the $CuO_2$-plane and $BiO_{1+\delta/2}$-layer hole concentrations with increasing $x$ are seen as enhancement in the relative intensities of the pre-edge peaks at ~528.3 and ~530.5 eV, respectively.

*PACS:* 61.10.Ht; 74.62.-c; 74.62.Dh; 74.72.Hs

The high-$T_c$ superconductive copper oxide, $M_mA_2Q_{n-1}Cu_nO_{m+2+2n\pm\delta}$ or $M$-$m2(n-1)n$, is believed to possess an antiferromagnetic insulating ground state related to its "undoped parent phase". By increasing the $CuO_2$-plane hole concentration the phase undergoes an insulator-metal transition and starts to show superconductivity with a transition temperature, $T_c$, that strongly depends on the concentration of induced holes. In the multi-layered structure of an $M$-$m2(n-1)n$ phase the superconductive $Q_{n-1}Cu_nO_{2n}$ block containing the $CuO_2$ plane(s) is sandwiched with two $AO$ layers and an $M_mO_{m\pm\delta}$ "charge reservoir" block with a layer sequence of $AO$-$CuO_2$-$(Q$-$CuO_2)_{n-1}$-$AO$-$(MO_{1\pm\delta/m})_m$.[1] Among the variety of known $M$-$m2(n-1)n$ phases ($M$ = *e.g.* Cu, Bi, Pb, Tl, Hg, Al, Ga, B; $m$ = 0 - 3; $A$ = *e.g.* Ba, Sr, La; $Q$ = *e.g.* Ca, rare-earth element $R$; $n$ = 1 - 9),[1] only a limited number of phases, *e.g.* $(La,Sr)_2CuO_{4\pm\delta}$ (0201), $CuBa_2RCu_2O_{7-\delta}$ (Cu-1212), $Bi_2Sr_2(R,Ca)Cu_2O_{8+\delta}$ (Bi-2212) and $(Tl_{0.5}Pb_{0.5})Sr_2(R,Ca)Cu_2O_{7\pm\delta}$ ((Tl,Pb)-1212),[2-6] allow us to experimentally observe the actual appearance of superconductivity adjacent to the insulator-metal boundary. This is because many of these phases are structurally rather weak to sustain doping within a sufficiently wide range. Another difficulty arises from the fact that no universal experimental tool to accurately probe the local $CuO_2$-plane hole concentration in the multi-layered copper-oxide superconductor has been realized yet.[1] Thus, for instance, the threshold $CuO_2$-plane hole concentration for the appearance of superconductivity has been established only for the simplest case, *i.e.*, for an $m$ = 0, $n$ = 1 system, $(La,Sr)_2CuO_{4\pm\delta}$, at 0.05 - 0.06.[2,3] Here we report the layer-specific hole concentrations in an $m$ = 2, $n$ = 2 system, $Bi_2Sr_2(Y_{1-x}Ca_x)Cu_2O_{8+\delta}$, within the whole $Ca^{II}$-for-$Y^{III}$ substitution range, *i.e.*, from an undoped insulating state ($x$ = 0) to a slightly overdoped state ($x$ = 1), as probed by two independent experimental techniques: x-ray absorption near-edge structure (XANES) spectroscopy and coulometric redox titration. The two techniques - a direct physical technique and an indirect but highly precise wet-chemical technique - are found to reveal highly consistent values for the actual $CuO_2$-plane hole concentration.

Since for an $M$-$m2(n-1)n$ phase with $n$ = 2 all the $CuO_2$ planes are equivalent, the $CuO_2$-plane hole concentration, $p(CuO_2)$, that is related to the nominal valence of copper[*], $V(Cu)$, according to $p(CuO_2) = V(Cu) - 2$, can be calculated for these phases from the stoichiometry of the phase when both the exact oxygen content and the valences of the other metals than copper are accurately established. In the case of the Bi-2212 phase, the analytical difficulties arise from the fact that, besides copper, bismuth may also exhibit mixed valence states. Distinguishing the individual valences of Cu and Bi is possible by means of a wet-chemical redox analysis method based on the selective reduction of $Bi^V$ (*i.e.* pentavalent Bi in the solid structure[**]) by $Fe^{2+}$ ions in an acidic solution and a subsequent electrochemical determination of the remaining amount of $Fe^{2+}$ ions by anodic oxidation.[1,7] The analysis can yield valence values with a high precision of ±0.01, but a critical question has remained to be addressed to the solution-based redox methods in general: how well are the solid-state characteristics, *i.e.*, the fine-distribution of electrons that applies when the atoms are arranged into the crystal lattice, maintained upon dissolving the material? Here the importance of searching for novel approaches and applying simultaneously various characterization methods is emphasized. XANES spectroscopy provides us with an ideal probe for the local concentration of holes as the x-ray absorption spectrum is determined by electronic transitions from a selected atomic core level to the unoccupied electronic states near the Fermi level.

For this study, a series of $Bi_2Sr_2(Y_{1-x}Ca_x)Cu_2O_{8+\delta}$ samples with $x = 0 - 1$ was prepared by solid-state reaction from stoichiometric mixtures of high-purity powders of $Bi_2O_3$, $SrCO_3$, $Y_2O_3$, $CaCO_3$ and $CuO$.[8] The mixed powders were first calcined in air at 770 - 830 °C for 12 hours and then sintered at 870 - 930 °C for 42 hours. Note that, the higher the Y content was, the higher synthesis temperature was required. The phase purity of the samples was checked by x-ray diffraction (XRD) measurements (Philips: PW 1830; Cu $K_\alpha$ radiation). The unit cell parameters were refined from the XRD data using a Rietveld refinement program, FULLPROF. The samples were further characterized for superconductivity properties by a SQUID magnetometer (Quantum Design: MPMS-5S). The $T_c$ values were taken as onset temperatures of the diamagnetic signal from the χ - *versus* - T curves measured from room temperature down to 5 K in a field-cooling mode under a magnetic field of 10 Oe.

The accurate oxygen contents were determined by coulometric $Cu^+/Cu^{2+}$ redox titration.[1,7] This experiment yields the total amount of high-valent copper and bismuth species, *i.e.*, $Cu^{III}$ and $Bi^V$, and thus the oxygen content of the sample. Upon dissolving the sample in 1 M HCl containing a known amount of $Cu^+$ ions both $Cu^{III}$ and $Bi^V$ oxidize $Cu^+$ to $Cu^{2+}$ according to reactions,

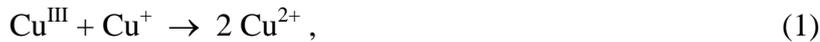
$$Cu^{III} + Cu^+ \rightarrow 2\ Cu^{2+}, \qquad (1)$$
and
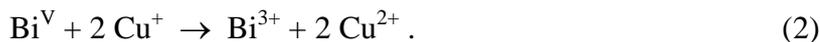
$$Bi^V + 2\ Cu^+ \rightarrow Bi^{3+} + 2\ Cu^{2+}. \qquad (2)$$

Once the reactions given by Eqs. (1) and (2) are completed the amount of remaining $Cu^+$ ions is accurately analyzed through coulometric titration, *i.e.*, anodic oxidation, as follows:

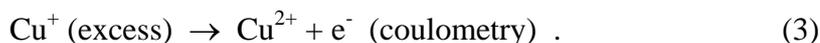
$$Cu^+\ (excess) \rightarrow Cu^{2+} + e^-\ (coulometry). \qquad (3)$$

From the amount of electrons produced in Eq. (3), the value of δ is calculated.[1,7]

The value of Bi valence was determined with another redox experiment, *i.e.*, $Fe^{2+}/Fe^{3+}$ coulometric titration.[1,7] This experiment allows us to detect selectively the amount of $Bi^V$ in the presence of $Cu^{III}$. The sample is dissolved in 1 M HCl containing a known amount of $Fe^{2+}$ ions. Pentavalent Bi reacts completely with $Fe^{2+}$ ions according to:

$$Bi^V + 2\ Fe^{2+} \rightarrow Bi^{3+} + 2\ Fe^{3+} \ . \qquad (4)$$

The valence of Bi, $V(Bi)_{tit}$, is obtained by analyzing the amount of $Fe^{2+}$ ions that did not participate in reaction given by Eq. (4) through anodic oxidation:

$$Fe^{2+}\ (excess) \rightarrow Fe^{3+} + e^-\ (coulometry)\ . \qquad (5)$$

Note that for $Cu^{III}$ reaction with water, *i.e.*,

$$4\ Cu^{III} + 2\ H_2O \rightarrow 4\ Cu^{2+} + O_2 + 4\ H^+\ , \qquad (6)$$

is more preferable than that with $Fe^{2+}$ ions, which prevents $Cu^{III}$ from interfering the determination of the valence of bismuth. The value of the Cu valence, $V(Cu)_{tit}$, can be calculated from the results of the oxygen content and Bi valence analyses, *i.e.*, values of δ and $V(Bi)_{tit}$, taking into account the cation stoichiometry of the phase.

The both redox experiments, *i.e.*, those described by Eqs. (1) - (3) and Eqs. (4) - (5), were carried out at room temperature under a flowing argon atmosphere. The 1 M-HCl cell solution was freed from dissolved oxygen by bubbling argon gas through it and the initial redox power of the cell was standardized by performing each time a pre-titration with a small amount of the corresponding reducing agent. As sources of the $Cu^+$ and $Fe^{2+}$ ions, $Cu_2O$ and $FeCl_2*4H_2O$, respectively, were used. Before the actual analyses, blank titrations were carried out to check the $Cu^+$ and $Fe^{2+}$ contents of these reductants. Each redox experiment was repeated at the minimum of five times to reveal the oxygen content and valence values with a reproducibility of less than ±0.01. The coulometric titration of $Cu^+$ ions [*cf*. Eq. (3)] was performed at a constant current of 5 mA until the potential of the AgCl/Ag indicator electrode reached 980 mV. The corresponding values in the case of the anodic oxidation of $Fe^{2+}$ ions [*cf*. Eq. (5)] were 3 mA and 820 mV.

The x-ray absorption experiments were carried out on the 6-m High-Energy Spherical Grating Monochromator (HSGM) beam-line at Synchrotron Radiation Research Center (SRRC) in Hsinchu, Taiwan. Both O *K*-edge and Cu $L_{2,3}$-edge XANES spectra were collected and the measurements were performed at room temperature. The powder samples were attached by conducting tape, and then put into an ultra high vacuum chamber (~$10^{-9}$ torr) in order to avoid surface contamination. The x-ray-fluorescence-yield spectra were recorded from the samples using a microchannel-plate (MCP) detector system consisting of a dual set of MCPs with an electrically isolated grid mounted in front of them. The grid was set to a voltage of 100 V, the front of the MCPs to -2000 V and the rear to -200 V. The grid bias ensured that positive ions did not enter the detector, while the MCP bias ensured that no electrons were detected. The detector was located parallel to the sample surface at a distance of ~2 cm. Photons were incident at an angle of 45° in respect to the sample normal. The incident photon flux ($I_0$) was monitored simultaneously by a Ni mesh located after the exit slit of the monochromator. All the absorption measurements were normalized to $I_0$. The photon energies were calibrated with an accuracy of 0.1 eV using the O $K$-edge absorption peak at 530.1 eV and the Cu $L_3$ white line at 931.2 eV of a CuO reference. The monochromator resolution was set to ~0.22 eV and ~0.45 eV at the O $K$ (1$s$) and Cu $L_{2,3}$ (2$p$) absorption edges, respectively. The x-ray-fluorescence-yield spectroscopy method applied is bulk-sensitive, the probing depth being 1000 - 5000 Å.

Judging from the x-ray-diffraction data, the synthesized $Bi_2Sr_2(Y_{1-x}Ca_x)Cu_2O_{8+\delta}$ samples are of single phase in the whole compositional range of $0 \leq x \leq 1$. From the wet-chemical redox analysis a continuous decrease from 0.52 to 0.25 in the amount of excess oxygen, $\delta$, is revealed upon increasing Ca content, $x$, from 0 to 1, as shown in Fig. 1. The obtained values for $\delta$ are essentially identical to those (ranging from 0.51 to 0.23) previously reported for air-synthesized $Bi_2Sr_2(Y_{1-x}Ca_x)Cu_2O_{8+\delta}$ samples based on iodometric titration.[9] As expected, with increasing $x$ the $c$-axis parameter of the unit cell increases, while both $a$- and $b$-axis parameters decrease (not shown). The expansion of the unit cell along the $c$ axis as $x$ increases is related to the fact that the ionic radius of $Ca^{II}$ is larger than that of $Y^{III}$. The contraction of the unit cell in the $a,b$-plane direction with increasing $x$ is due to an increase in the overall hole concentration, $p_{tot}$.

We calculate $p_{tot}$ for each sample from the known cation stoichiometry and the analyzed oxygen content as the amount of holes *per* half formula unit, *i.e.*, the sum of the $CuO_2$-plane and $BiO_{1+\delta/2}$-layer hole concentrations. In Fig. 1, $p_{tot}$ is given against $x$. As the $Ca^{II}$-for-$Y^{III}$ substitution proceeds, $p_{tot}$ increases monotonically even though $\delta$ decreases. From Fig. 1, with increasing $p_{tot}$ the hole concentration of the $CuO_2$ plane, $p(CuO_2)_{tit}$, calculated from $V(Cu)_{tit}$ as $p(CuO_2)_{tit} = V(Cu)_{tit} - 2$, increases from 0.02 to 0.12 when $x$ ranges from 0 to 1 in $Bi_2Sr_2(Y_{1-x}Ca_x)Cu_2O_{8+\delta(x)}$. Within the whole substitution range, the value of $p(CuO_2)_{tit}$ is essentially lower than that of $p_{tot}$, owing to the fact that upon oxidizing the phase some part of holes goes into the $Bi_2O_{2+\delta}$ charge-reservoir block. The value of the concentration of holes in one $BiO_{1+\delta/2}$ layer given by $p(BiO_{1+\delta/2})_{tit} = V(Bi)_{tit} - 3$ increases with increasing $x$ from 0.00 to 0.13, despite the fact that the amount of excess oxygen, $\delta$, in the $Bi_2O_{2+\delta}$ block decreases (Fig. 1).

The increase in the $CuO_2$-plane hole concentration with Ca substitution is clearly revealed from the O $K$-edge XANES data. Figure 2 displays the O $K$-edge XANES spectra obtained for the samples in the energy range of 525 to 555 eV. With increasing $x$, a pre-edge peak develops around 528.3 eV. This peak was previously ascribed to holes in the singlet band formed on $p$-type doping of the $CuO_2$ plane in Bi-2212, $i.e.$, transition from $3d^9L$ to $O1s3d^9$ ($L$ denotes a hole in an $O2p_{xy}$ orbital).[10-13] A similar feature has been established for many other $p$-type doped superconductive copper oxides.[14-16] The peak at ~530.5 eV is due to the wide antibonding $Bi6p_{x,y,z}$-O and $O2p_{x,y}$-$O2p_z$ band, while the shoulder at ~529.5 eV, that is most clearly seen for the $x = 0$ sample, arises from a transition into $O2p$ states hybridized with the upper Hubbard band (UHB).[10-13] The O $K$-edge spectral features are analyzed by fitting Gaussian functions to the three pre-edge peaks. Before the fitting the spectra are first normalized to have the same spectral weight in the energy range of 535 to 555 eV, and then multiplied by the exact oxygen content of the sample as determined by wet-chemical analysis.[17] The latter operation is not highly meaningful but rather theoretical as it affects the final results less than 3 %. In Fig. 3, the integrated intensities of the peaks at ~528.3, ~529.5 and ~530.5 eV, $i.e.$, $I_{528.3}$, $I_{529.5}$ and $I_{530.5}$, respectively, are plotted against the Ca content, $x$. The continuous increase in $I_{528.3}$ with increasing $x$ directly reflects the increase in the $CuO_2$-plane hole concentration. The absorption energy of the "528.3-eV peak" slightly decreases with increasing intensity. This demonstrates that the Fermi level moves to a lower energy when the hole concentration within the $CuO_2$ plane increases. At the same time, the density of states in UHB is diminished, $i.e.$, $I_{529.5}$ decreases as observed in Fig. 3. The increase in $I_{530.5}$ with increasing Ca substitution level may be considered as an indication of increase in the valence value of bismuth. From Fig. 3, the magnitude of the increase in the spectral weight is roughly the same for the $I_{528.3}$ and $I_{530.5}$ peaks, suggesting that the valences of Cu and Bi increase with quite similar rates with increasing Ca substitution level. This is highly consistent with the redox titration data given in Fig. 1 for the Cu and Bi valence values.

Quantitative estimations for the $CuO_2$-plane hole concentration are obtained from the Cu $L_{2,3}$-edge XANES spectra. The Cu $L_{2,3}$-edge spectra measured for the samples in the energy range of 925 to 955 eV are shown in Fig. 4. For the $x = 1.0$ sample, the spectrum exhibits two narrow peaks centered at ~931.2 and ~951.2 eV. These peaks are due to divalent copper states, $i.e.$, transitions from the $Cu(2p_{3/2,1/2})3d^9$-$O2p^6$ ground-state configuration into the $Cu(2p_{3/2,1/2})^-$

$^{-1}3d^{10}$-O$2p^6$ excited state, where $(2p_{3/2,1/2})^{-1}$ denotes a $2p_{3/2}$ or $2p_{1/2}$ hole.[18] Oxidation of copper beyond the divalent state is seen as shoulders on the high-energy side of these peaks. Such shoulders, first observed for fully-oxygenated CuBa$_2$YCu$_2$O$_{7-\delta}$[19] and later for various Bi-2212 samples[20-23], are interpreted as transitions from the Cu($2p_{3/2,1/2}$)$3d^9L$ ground state into the Cu($2p_{3/2,1/2}$)$^{-1}3d^{10}L$ excited state, where $L$ denotes a ligand hole in the O$2p$ orbital, i.e., being due to Cu$^{III}$. With increasing $x$, both the absorption peaks become more asymmetric, as the $2p$ hole concentration on the oxygen site increases leading to an increase in the intensity of the high-energy shoulders. For each sample, the spectrum is analyzed by fitting the $L_3$ peak (that is more intense than the $L_2$ peak) and its shoulder with Gaussian functions. The integrated intensity of the shoulder [$I$(Cu$^{III}$)] is normalized against the total spectral weight in the $L_3$ area below 935 eV, i.e., the sum of integrated intensity of the main peak [$I$(Cu$^{II}$)] and that of the shoulder itself. The normalized intensity of the shoulder, i.e., $I$(Cu$^{III}$) / [$I$(Cu$^{III}$) + $I$(Cu$^{III}$)], gives the ratio of the amount of Cu$^{III}$ to the total amount of Cu$^{II}$ and Cu$^{III}$, thus being nothing but a direct estimation for the hole concentration of the CuO$_2$ plane, $p$(CuO$_2$)$_{XAS}$. In Fig. 5, the thus obtained $p$(CuO$_2$)$_{XAS}$ values are plotted against $x$, together with the CuO$_2$-plane hole concentration values estimated based on the coulometric redox analysis. Also given are the values of $p_{tot}$ for reference. As $x$ increases from 0 to 1, $p$(CuO$_2$)$_{XAS}$ increases from 0.04 to 0.15. The value of 0.15 obtained for the $x = 1$ sample (Bi$_2$Sr$_2$CaCu$_2$O$_{8.52}$) is exactly the same as previously reported for similar samples based on Cu $L_3$-edge XANES analysis.[21,23] For the whole sample series, the $p$(CuO$_2$)$_{XAS}$ values ranging from 0.04 to 0.15 are also in good agreement with those of $p$(CuO$_2$)$_{tit}$ ranging from 0.02 to 0.12, within the error limits estimated for the two analysis techniques, i.e., ±0.01 for $p$(CuO$_2$)$_{tit}$ and ±0.02 for $p$(CuO$_2$)$_{XAS}$. Moreover, both the analyses clearly reveal that the actual $p$(CuO$_2$) values are essentially lower than the $p_{tot}$'s.

In Fig. 6, the relationship between the value of $T_c$ and the CuO$_2$-plane hole concentration, $p$(CuO$_2$), taken as an average of $p$(CuO$_2$)$_{tit}$ and $p$(CuO$_2$)$_{XAS}$, is shown. Superconductivity appears in the Bi$_2$Sr$_2$(Y$_{1-x}$Ca$_x$)Cu$_2$O$_{8+\delta(x)}$ system for $x = 0.4 - 0.5$ with increasing Ca content. From Fig. 6, the threshold CuO$_2$-plane hole concentration for the appearance of superconductivity can be established at 0.06 ± 0.01. With increasing Ca$^{II}$-for-Y$^{III}$ substitution level, $T_c$ increases up to $x \approx 0.8$ such that the maximum $T_c$ of ~90 K is observed at $p$(CuO$_2$) $\approx$ 0.12. The $x = 1$ sample with $p$(CuO$_2$) $\approx$ 0.14 is considered to be already slightly overdoped. In terms of the appearance of superconductivity the present threshold $p$(CuO$_2$) value of ~0.06 coincides with that established for the (La,Sr)$_2$CuO$_{4\pm\delta}$ system.[2,3] On the other hand, the $p$(CuO$_2$)$_{opt}$ value of ~0.12 revealed for the Bi$_2$Sr$_2$(Y$_{0.2}$Ca$_{0.8}$)Cu$_2$O$_{8.30}$ sample with the highest $T_c$ is somewhat low if one expects a value close to 0.16.[24] Here we like to note, however, that $p(CuO_2)_{opt}$ being at ~0.16 has been experimentally established only for $(La,Sr)_2CuO_{4\pm\delta}$. The Hg-based single-$CuO_2$-plane copper oxide, $HgBa_2CuO_{4+\delta}$, is another phase for which determination of the actual $CuO_2$-plane hole concentration should be straightforward. Nevertheless, no direct evidence pointing out at $p(CuO_2)_{opt}$ being at 0.16 has been presented. For an optimally doped $HgBa_2CuO_{4+\delta}$ sample, a value of 0.18 was revealed based on O $K$-edge XANES analysis.[25] Estimations based on the amount of excess oxygen in $HgBa_2CuO_{4+\delta}$ typically result in even higher $p(CuO_2)_{opt}$ values if integer valence values of II and –II, respectively, are assumed for Hg and O atoms in the $HgO_\delta$ charge reservoir.[26] For optimally doped $CuBa_2YCu_2O_{7-\delta}$ O $K$-edge XANES data revealed a $CuO_2$-plane hole concentration of 0.20,[27] while quantitative analysis of reflection intensities of convergent-beam electron diffraction data ended up to a value of 0.25.[28] For the three-$CuO_2$-plane phases the question on the optimum $CuO_2$-plane hole concentration is rather complicated and far from understood yet.[1] Against this discussion the present result of the $p(CuO_2)_{opt}$ value being at ~0.12 in the $Bi_2Sr_2(Y_{1-x}Ca_x)Cu_2O_{8+\delta}$ system suggests that the precise mechanism behind the $T_c$ degradation in the so-called overdoped region may not be totally equivalent among the various high-$T_c$ superconductive systems.

In conclusion, utilizing two independent analytical techniques, *i.e.*, XANES spectroscopy and wet-chemical redox analysis, we have here unambiguously revealed that for $Bi_2Sr_2(Y_{1-x}Ca_x)Cu_2O_{8+\delta}$ samples sintered in air not only the oxygen content and the valence of Cu ($CuO_2$-plane hole concentration) but also the valence of Bi (charge-reservoir hole concentration) change gradually as the $Ca^{II}$-for-$Y^{III}$ substitution proceeds. Moreover, excellent quantitative agreements have been demonstrated in the magnitude of the valence values estimated through the two techniques.

Dr. P. Nachimuthu is acknowledged for his help in XANES experiments. This work was supported by Academy of Finland (Decision No. 46039) and by a Grant-in-Aid for Scientific Research (contract No. 11305002) from the Ministry of Education, Science and Culture of Japan.

---

\* In covalently bonded oxide layers such as the $CuO_2$ plane and the $BiO_{1+\delta/2}$ layer in $Bi_2Sr_2(Y_{1-x}Ca_x)Cu_2O_{8+\delta}$ the nominal cation valence may be calculated assuming a valence of –II for oxygen. Note also that the presently applied analysis techniques do not really distinguish the

valence of Cu/Bi from that of the surrounding oxygen within the same plane/layer, *i.e.*, they rather detect the overall concentration of holes within the plane/layer.

** To indicate the oxidation state/valence of a species in solid matrix, Roman numerals are used (*e.g.* $Cu^{III}$), while the charges of ions in solution are indicated by Arabic numerals (*e.g.* $Cu^{2+}$).

**Figure Captions**

**Fig. 1.** The amount of excess ogygen, $\delta$, the overall hole content *per* half formula unit, $p_{tot}$, the $CuO_2$-plane hole concentration, $p(CuO_2)_{tit}$, and the $BiO_{1+\delta/2}$-layer hole concentration, $p(BiO_{1+\delta/2})_{tit}$, as established by coulometric redox analysis with increasing Ca-substitution level, $x$, for the $Bi_2Sr_2(Y_{1-x}Ca_x)Cu_2O_{8+\delta}$ samples. As $x$ increases not only $p_{tot}$ and $p(CuO_2)_{tit}$ but also $p(BiO_{1+\delta/2})_{tit}$ increase, whereas $\delta$ decreases.

**Fig. 2.** O $K$-edge XANES spectra for the $Bi_2Sr_2(Y_{1-x}Ca_x)Cu_2O_{8+\delta}$ samples in the energy range of 525 - 555 eV.

**Fig. 3.** Integrated intensities of the pre-edge peaks in the O $K$-edge XANES spectra at ~528.3 ~529.5 and ~530.5 eV, *i.e.* $I_{528.3}$, $I_{529.5}$ and $I_{530.5}$, with respect to the Ca-substitution level, $x$, for the $Bi_2Sr_2(Y_{1-x}Ca_x)Cu_2O_{8+\delta}$ samples. The changes of $I_{528.3}$ and $I_{530.5}$, respectively, are parallel to $p(CuO_2)_{tit}$ and $p(BiO_{1+\delta/2})_{tit}$ given in Fig. 1.

**Fig. 4.** Cu $L_{2,3}$-edge XANES spectra for the $Bi_2Sr_2(Y_{1-x}Ca_x)Cu_2O_{8+\delta}$ samples in the energy range of 925 - 965 eV. The peak at ~932 eV and the shoulder at ~933.5 eV correspond to $Cu^{II}$ and $Cu^{III}$ (*i.e.* the hole state in the $CuO_2$ plane), respectively.

**Fig. 5.** The $CuO_2$-plane hole concentration, $p(CuO_2)_{XAS}$, as calculated from the fitted Cu $L_3$-edge XANES data with respect to the Ca-substitution level, $x$, for the $Bi_2Sr_2(Y_{1-x}Ca_x)Cu_2O_{8+\delta}$ samples. Note that $p(CuO_2)_{XAS}$ are in good agreement with $p(CuO_2)_{tit}$ (given in Fig. 1) within the experimental error limits. Also note that both $p(CuO_2)_{XAS}$ and $p(CuO_2)_{tit}$ are always lower than $p_{tot}$.

**Fig. 6.** The relationship between $T_c$ and the $CuO_2$-plane hole concentration, $p(CuO_2)$, in the $Bi_2Sr_2(Y_{1-x}Ca_x)Cu_2O_{8+\delta}$ system. Note that, $p(CuO_2)$ is taken as an average of the values determined for the $CuO_2$-plane hole concentration by coulometric redox analysis [$p(CuO_2)_{tit}$] and by Cu $L_3$-edge XANES spectroscopy [$p(CuO_2)_{XAS}$]. The threshold hole concentration for the appearance of superconductivity is seen at $p(CuO_2) \approx 0.06$.

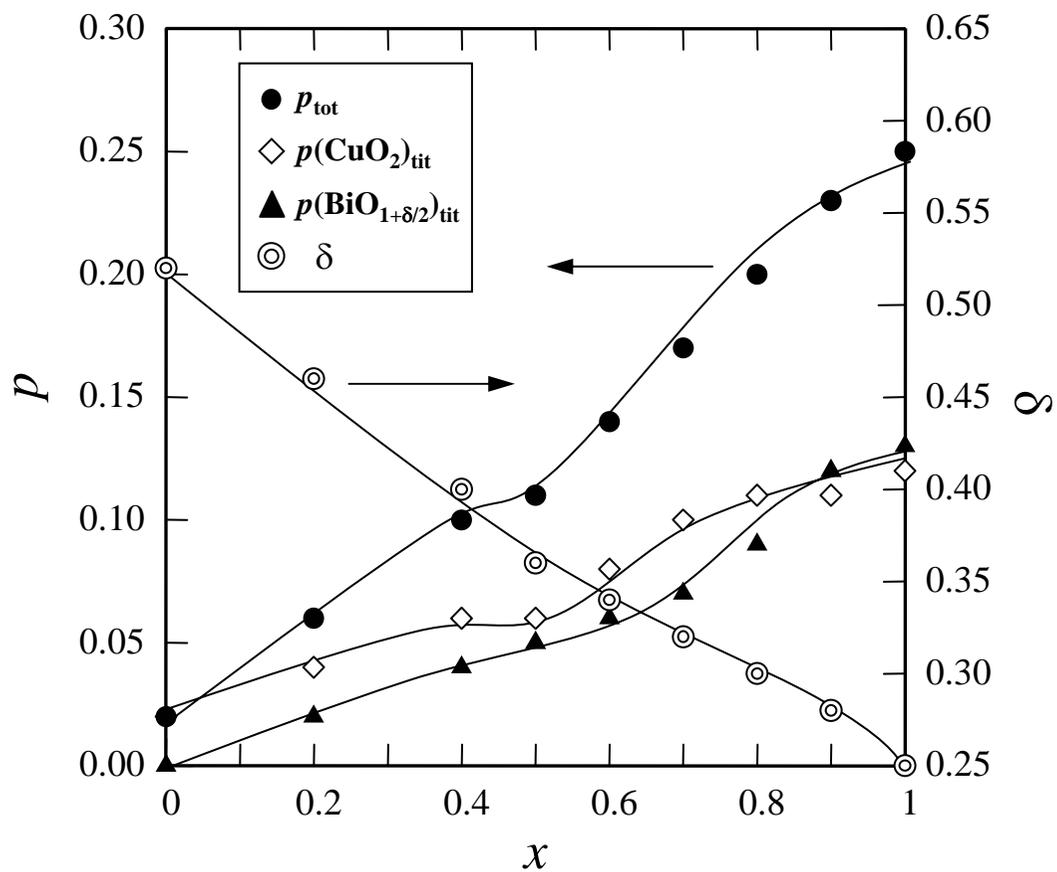

Karppinen *et al*: Fig. 1.

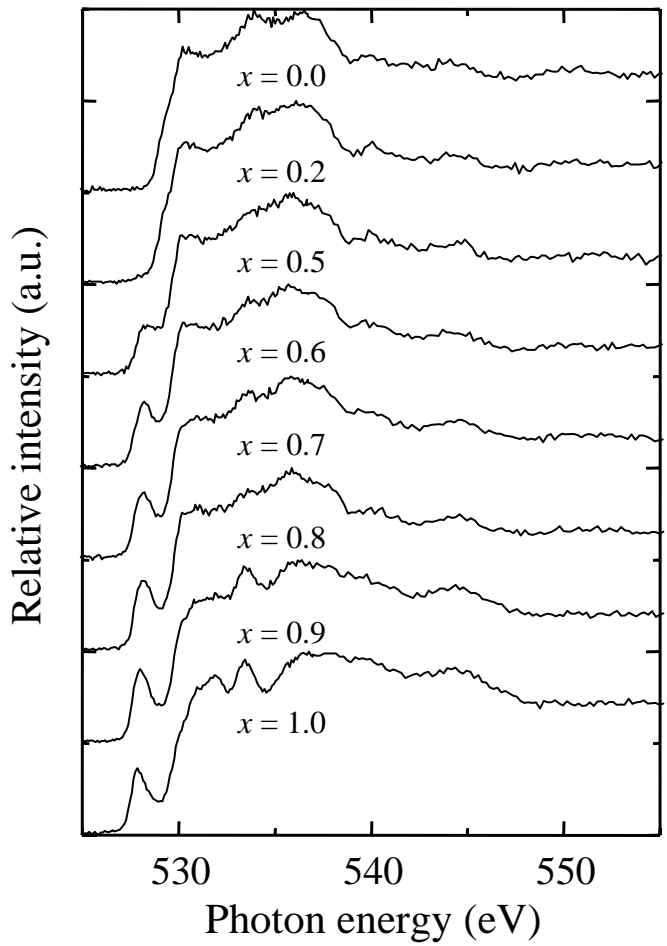

Karppinen *et al*: Fig. 2.

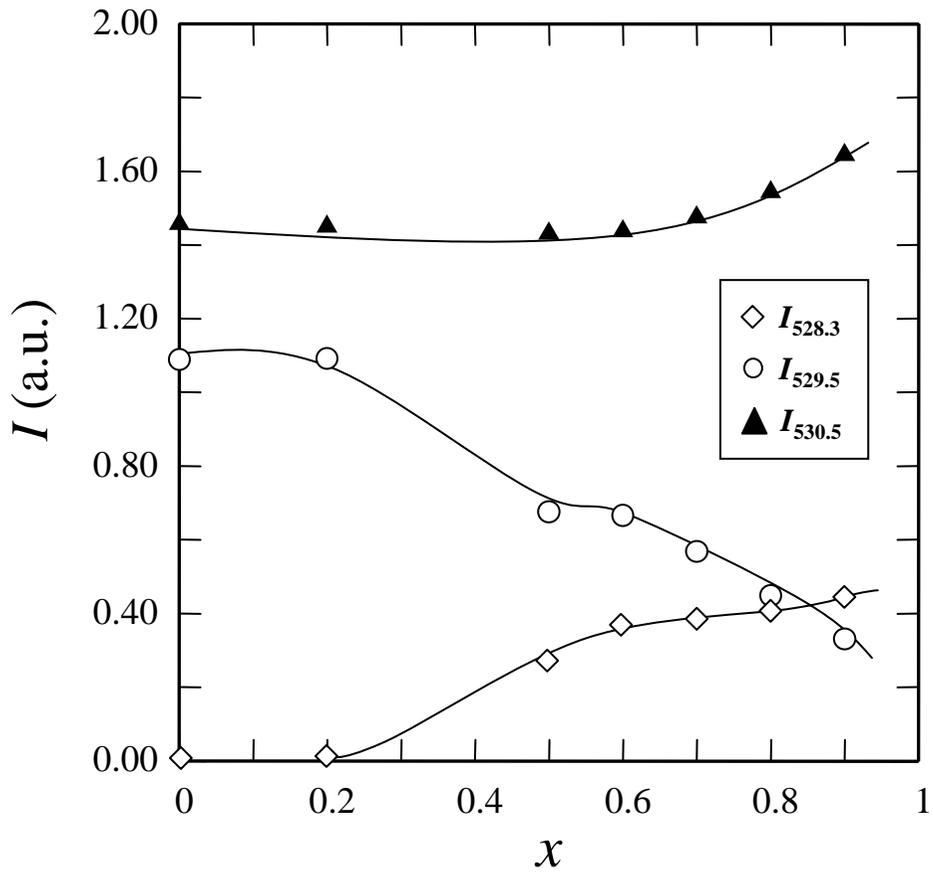

Karppinen *et al*: Fig. 3.

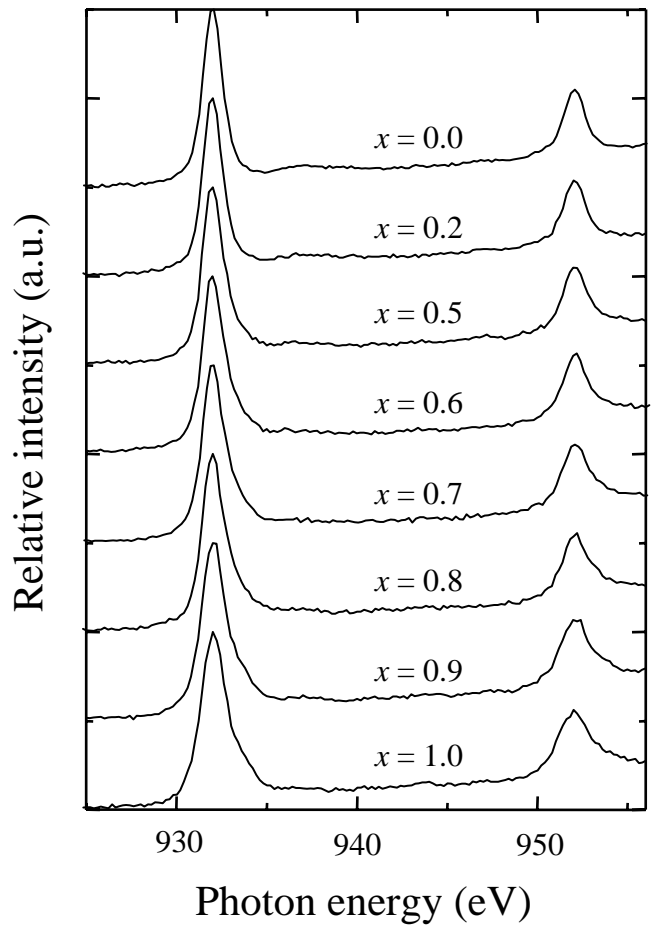

Karppinen *et al*: Fig. 4.

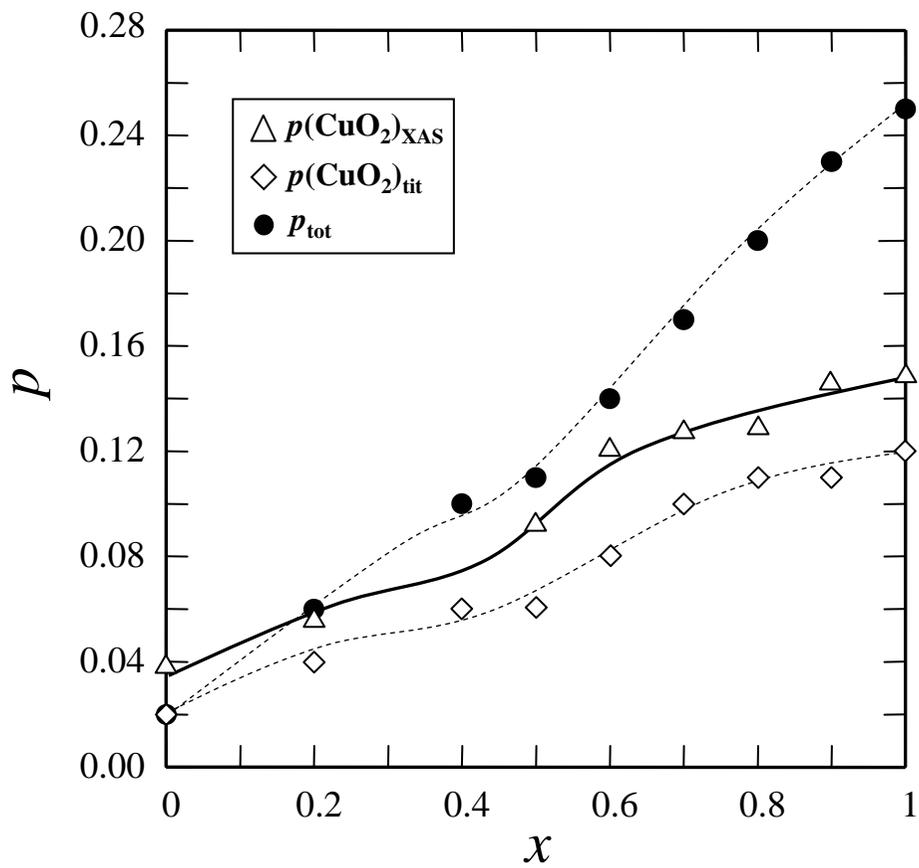

Karppinen *et al*: Fig. 5.

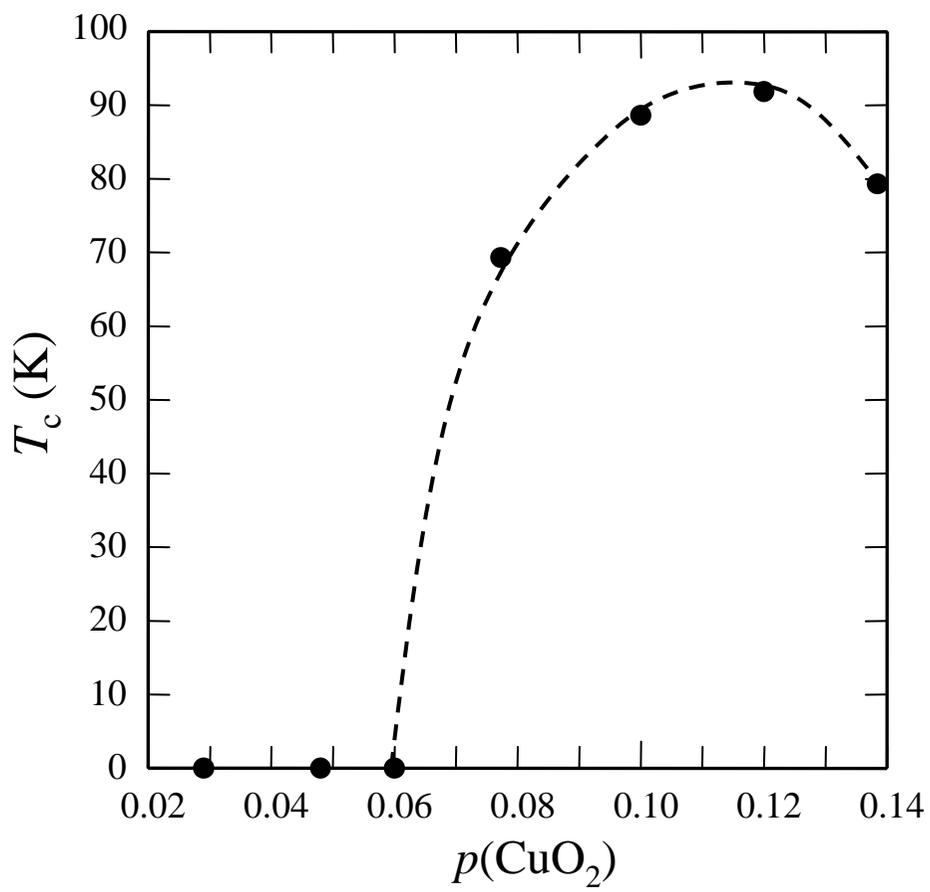

Karppinen *et al*: Fig. 6.